\newcommand{\dfd}{{\rm d}}
\newcommand{\vecr}{\mbox{\boldmath$r$}}

\documentclass[12pt,a4paper]{article}
\usepackage[]{graphicx}\usepackage[]{epsfig}
\title{Electrostatic interaction energies of homogeneous cubic charge distributions}
\author{Hanno Ess\'en\\Department of Mechanics \\Royal Institute of Technology  \\
S-100 44 Stockholm, Sweden}
\begin{document}
\maketitle
\begin{abstract}
The starting point is the problem of finding the interaction energy of two
coinciding homogeneous cubic charge distributions. The brute force method of
subdividing the cube into $N^3$ sub-cubes and doing the sums results in slow
convergence because of the Coulomb singularity. Using symmetry and algebra the
Coulomb singularities can be eliminated. This leads to an accurate numerical
algorithm as well as an interesting exact result relating the desired interaction
energy to three other interaction energies, namely those of cubes touching each
other at a face, at an edge, and at a corner, respectively. As an application a
simple model illustrating Wigner crystallization is presented.
\end{abstract}
\newpage
\section{Introduction}
There are still many interesting problems involving the
electrostatics of cubic geometries. These have to do with cubic
ionic crystals \cite{essen&nordmark,moggia}, with the force and
potential from cubic charge and mass distributions
\cite{waldvogel,chen&cook,hummer,seidov}, and with the electric
capacitance of the cube \cite{reitan,hwang}. Here we will discuss
the evaluation of the electrostatic interaction energy of two
coinciding homogeneous cubic charge distributions. For unit charge
distributions in unit cubes this energy is given by,
\begin{equation}\label{eq.C.def} C =  \int_{V_1} \int_{V_2} \frac{\dfd V_1\, \dfd
V_2}{|\vecr_1- \vecr_2|},
\end{equation}
where, $\vecr_a=(x_a, y_a, z_a)$,
and
\begin{equation}\label{eq.cubes.defs} V_{a} = \{ (x_a, y_a, z_a) ;\; 0\!
<x_a\!<\! 1,\; 0\! <y_a\! <\! 1,\; 0\! <z_a\! <\! 1 \,  \}, \hskip
0.6cma=1,2.
\end{equation}
This integral arises naturally in the free electron gas theory of conduction
electrons in metals, see Raimes \cite{raimes}. Its actual value is usually not
needed for most applications of that theory. In an extension of the theory by
Ess\'en \cite{essen95}, however, it is needed. Another application of this type of
integral will be given below. The value of $C$ can be calculated exactly. Put,
\begin{equation}
\phi_C (\vecr) = \int_{V_2} \frac{\dfd V_2}{|\vecr- \vecr_2|},
\end{equation}
for the electrostatic potential energy from a homogeneous cubic
charge distribution. This potential has been discussed by
Waldvogel \cite{waldvogel}, by Hummer \cite{hummer}, and by Seidov
and Skvirsky \cite{seidov}. Using it we can write (\ref{eq.C.def})
in the form,
\begin{equation}\label{eq.C.def.with.phi} C =  \int_{V_1} \phi_C (\vecr_1)\,
\dfd V_1,
\end{equation}and this makes it possible to find the analytical
expression,
\begin{equation}\label{eq.C.analytical.expression}
C =-2\left\{
\frac{2\sqrt{3}-\sqrt{2}-1}{5} +\frac{\pi}{3}
+\ln\left[(\sqrt{2}-1)(2-\sqrt{3})\right]  \right\},
\end{equation}
(Seidov and Skvirsky \cite{seidov}). This evaluates
to,
\begin{equation}\label{eq.C.20.dig}C \approx
1.8823126443896601600
\end{equation}
using twenty digits. Another expression for $C$ in terms of a one dimensional
integral has been derived by Ess\'en and Nordmark \cite{essen&nordmark}.

If we displace one of the cubes in (\ref{eq.cubes.defs}) one unit
along the $x$-axis the integral (\ref{eq.C.def}) changes into an
integral for the interaction energy of cubes with one face
touching (see Fig.\ \ref{FIG1}). Let us call this integral $C_{\rm
f}$. If we displace one cube one unit along both the $x$ and the
$y$-axis we get the integral for cubes with an edge in common,
call it $C_{\rm e}$. If we finally displace one of the cubes one
unit along all three directions of space we get the integral for
cubes touching at one corner, call it $C_{\rm c}$. One of the
results found below then says that,
\begin{equation}\label{eq.result.for.the.four}
C = C_{\rm f} + C_{\rm e}+ \frac{1}{3}C_{\rm c}.
\end{equation}
This might be a new result.

Coulomb interaction energy integrals find one of their main
applications in Hartree and Hartree-Fock self-consistent field
studies of many electron systems, see for example Raimes
\cite{raimes}. As an application of the results of this paper we
use them for crude estimates of the energy of electrons moving in
a cubic background of smeared out positive charge. In particular
we compare the energies of delocalized electron states with those
of localized states. When the density is small the localized
states are found to have lower energy. This is the phenomenon of
Wigner crystallization \cite{wigner34,wigner38}.

\section{The brute force approach}
Consider two electrons of charge $e$ in a cubic box with edges of
length $L$. Assume that both electrons have constant charge
density,
\begin{equation}
\rho = e/L^3,
\end{equation}
in this box. The Coulomb, electrostatic, interaction energy of these charge
distributions is then:
\begin{equation}
C\, \frac{e^2}{L} = \left(\frac{e}{L^3}\right)^2 \int_{\vecr_1 \in V_1} \left(
\int_{\vecr_2 \in V_2} \frac{ \dfd V_2}{|\vecr_1 -\vecr_2|} \right) \dfd V_1  .
\end{equation}
Here $V_{a}, (a=1,2)$ denote the cubic boxes over which the
integration variables, $\vecr_a=(x_a, y_a, z_a)$, take their
values. We now introduce units so that $e = L = 1$. The integral
can then be expressed in the form,
\begin{equation}\label{eq.for.C.six.dim.explicit} C = \int_{x =0}^{x =1}
\!\int_{y =0}^{y =1} \! \int_{z =0}^{z =1} \! \int_{u =0}^{u =1} \!\int_{v =0}^{v
=1} \! \int_{w =0}^{w =1} \! \! \frac{\dfd x\, \dfd y\, \dfd z\, \dfd u\, \dfd v\,
\dfd w }{\sqrt{ (x\! -\! u)^2 +(y\! -\! v)^2 + (z\! -\! w)^2 } } ,
\end{equation}
which shows explicitly that this is a six-dimensional integral.

Nowadays we are spoilt by systems for doing mathematics by
computer. It is therefore tempting to try these systems whenever
some cumbersome integral arises, and frequently they do deliver
sensible answers. For the integral
(\ref{eq.for.C.six.dim.explicit}), however, those that I have
tried fail. Brute force can't handle the Coulomb singularity. Let
us see what happens if we start by dividing each cube into $N^3$
sub-cubes:
\begin{equation}
V_{aN}^{ijk} = \left\{ (x_a, y_a, z_a) ;\; \frac{i-1}{N} \! <x_a\! <\!
\frac{i}{N},\; \frac{j-1}{N} \! <y_a\! <\! \frac{j}{N},\; \frac{k-1}{N} \! <z_a\!
<\! \frac{k}{N} \right\},
\end{equation}
where the indices, $i, j,$ and $k$, run from 1 to $N$. Our
integral can then be written as the sum,
\begin{equation}
C = \sum_{ijk=1}^{N} \sum_{lmn=1}^{N} C^{lmn}_{N,ijk},
\end{equation}
over $N^6$ terms, integrals over pairs of sub-cubes,
\begin{equation}
C^{lmn}_{N,ijk} = \int_{V_{1N}^{ijk}} \int_{V_{2N}^{lmn}}
\frac{\dfd V_1\, \dfd V_2}{|\vecr_1- \vecr_2|} .
\end{equation}
For sufficiently large $N$ most integrals are over pairs of
spatially separated sub-cubes and can be easily approximated. This
leads to a brute force approach. Fairly large contributions
should, however, come from pairs of cubes that coincide or touch
since they are strongly affected by the singularity. Such an
approach is clearly clumsy.

\section{Removing the interior singularity}
The awkward singularity occurs only in the interior of those $N^3$
terms of this sum for which the integration sub-cubes are equal. If we thus write,
\begin{equation}\label{eq.two.sums}
C = \sum_{ijk=1}^{N}  C^{ijk}_{N,ijk}
+\sum_{ijk=1}^{N} {\sum_{lmn=1}^{N}}' C^{lmn}_{N,ijk} ,
\end{equation}
where the terms with all three indices the same ($i=l, j=m, k=n$)
are excluded in the double sum, we see that the interior
singularities occur in the first sum over coinciding sub-cubes.
But these integrals are all identical and equal to,
\begin{equation}
C_N = C^{111}_{N,111}
= \int_{V_{1N}^{111}} \int_{V_{2N}^{111}} \frac{\dfd V_1\, \dfd V_2}{|\vecr_1-
\vecr_2|}.
\end{equation}
From formula (\ref{eq.two.sums}) one thus
gets,
\begin{equation}\label{eq.simp.sums} C = N^3 C_N +
\sum_{ijk=1}^{N}{\sum_{lmn=1}^{N}}' C^{lmn}_{N,ijk}.
\end{equation}
Apart from being over a smaller cube, the integral $C_N$ is essentially like the
original integral. In fact one easily finds the scaling
property,
\begin{equation}\label{eq.scaling}
C = N^5 C_N .
\end{equation}
Using this equation (\ref{eq.simp.sums}) becomes,
\begin{equation}
C = \frac{C}{ N^2} +
\sum_{ijk=1}^{N} {\sum_{lmn=1}^{N}}'C^{lmn}_{N,ijk} .
\end{equation}
Solving for $C$
we thus finally have,
\begin{equation}\label{eq.no.sing.N}
C = \frac{N^2}{N^2 -1}
\sum_{ijk=1}^{N}{\sum_{lmn=1}^{N}}' C^{lmn}_{N,ijk}.
\end{equation}
Here the original integral with its singularity has been written as a sum of $N^6 -
N^3$ integrals without (interior) singularities.

\section{Approximating the non-diagonal integrals}
The non-singular integrals can be approximated by the
product of the two cubic volumes divided by the distance between their midpoints. A
simple calculation gives,
\begin{equation}C^{lmn}_{N,ijk} = \int_{V_{1N}^{ijk}}
\int_{V_{2N}^{lmn}} \frac{\dfd V_1\, \dfd V_2}{|\vecr_1- \vecr_2|}
\approx\frac{1}{N^5} \frac{1}{\sqrt{(i-l)^2 + (j-m)^2 + (k-n)^2}} .
\end{equation}
If we  introduce the notation, compare equation (\ref{eq.scaling}),
\begin{equation}C^{lmn}_{ijk} \equiv N^5
C^{lmn}_{N,ijk},
\end{equation}
we now have,
\begin{equation}\label{eq.no.sing} C =
\frac{1}{N^3 (N^2 -1)} \sum_{ijk=1}^{N}{\sum_{lmn=1}^{N}}'
C^{lmn}_{ijk}.
\end{equation}
Since the number of terms in the sum grows as $N^6$ it is of
interest to take advantage of symmetries to reduce it as much as
possible. Doing this we find that,
\begin{equation}\label{eq.I.sum.rewritten}
C = \frac{12}{N^3(N^2 -1)}
\left(\frac{N^2}{2} \sum_{l<i}^{N} C_{i11}^{l11}\label{eq.sum.algorithm}
+N\sum_{m<j}^{N} \sum_{n<k}^{N}C_{1jk}^{1mn}+ \frac{2}{3}
\sum_{l<i}^{N}\sum_{m<j}^{N}\sum_{n<k}^{N} C_{ijk}^{lmn} \right).
\end{equation}
is an alternative way of writing equation (\ref{eq.no.sing}). Putting,
\begin{equation}
\Delta^{lmn}_{ijk} \equiv \sqrt{(i-l)^2 +
(j-m)^2 + (k-n)^2} ,
\end{equation}
we have
that,
\begin{equation}\label{eq.basic.approximation} C^{lmn}_{ijk}
\approx1/\Delta^{lmn}_{ijk} ,
\end{equation}
assuming that $(i,j,k)\neq(l,m,n)$. If we put this into (\ref{eq.no.sing}), or
(\ref{eq.I.sum.rewritten}), we get,
\begin{eqnarray}\label{eq.brute.sum.I} &\displaystyle C \approx C_N^0
\equiv\frac{1}{N^3(N^2 -1)} \sum_{ijk=1}^{N} {\sum_{lmn=1}^{N}}'
\frac{1}{\Delta^{lmn}_{ijk}} =&\\ &\displaystyle\frac{12}{N^3(N^2 -1)} \left(
\frac{N^2}{2} \sum_{l<i}^{N}\frac{1}{\Delta_{i11}^{l11} } +N\sum_{m<j}^{N}
\sum_{n<k}^{N}\frac{1}{\Delta_{1jk}^{1mn}}+
\frac{2}{3}\sum_{l<i}^{N}\sum_{m<j}^{N}\sum_{n<k}^{N}\frac{1}{\Delta_{ijk}^{lmn}}
\right).&
\end{eqnarray}
The smaller the box, the smaller the error, so there is hope that this expression
will converge to the correct value of $C$ when $N$ goes to infinity, i.e.\ that
\begin{equation}\label{eq.limit.for.I}
C = \lim_{N\rightarrow \infty} C_N^0 .
\end{equation}
The approximation (\ref{eq.basic.approximation}) then immediately gives the
following estimate for $C$, when $N=2$,
\begin{equation}C \approx C_2^0 = 1 +
\frac{1}{\sqrt{2}} +\frac{1}{3}\frac{1}{\sqrt{3}} \approx
1.899556871,
\end{equation}
a value which turns out to be correct to two significant digits. This is encouraging
but the convergence for increasing $N$ is slow, see Table \ref{table.1st.approx}.
When $N=16$ the error is still $3\cdot 10^{-4}$. The exact value is from Eqs.\
(\ref{eq.C.analytical.expression}) and (\ref{eq.C.20.dig}).

\begin{table}
\begin{tabular}{|c|l|l|} \hline
$N$ & $C_N^0$ & $C_N^0 - C$ \\
\hline  $2$ &  $1.899556871$ & $2\cdot 10^{-2}$ \\
\hline  $4$ &  $1.887296187$ & $5\cdot 10^{-3}$ \\
\hline  $8$ &  $1.883654361$ & $1\cdot 10^{-3}$ \\
\hline $16$ & $1.882660569$ & $3\cdot 10^{-4}$  \\
\hline $\vdots$&  $\vdots$    & $\vdots$        \\
\hline $\infty$&$1.882312644$ & $0$             \\
\hline
\end{tabular}
\caption[firstapprox]{\small \label{table.1st.approx} This table illustrates the
slow convergence of $C_N^0$. Note that computation time goes as
$N^6$.}
\end{table}

\section{Removing the remaining singularity}

\begin{figure}[h] \centering\includegraphics[width=200pt]{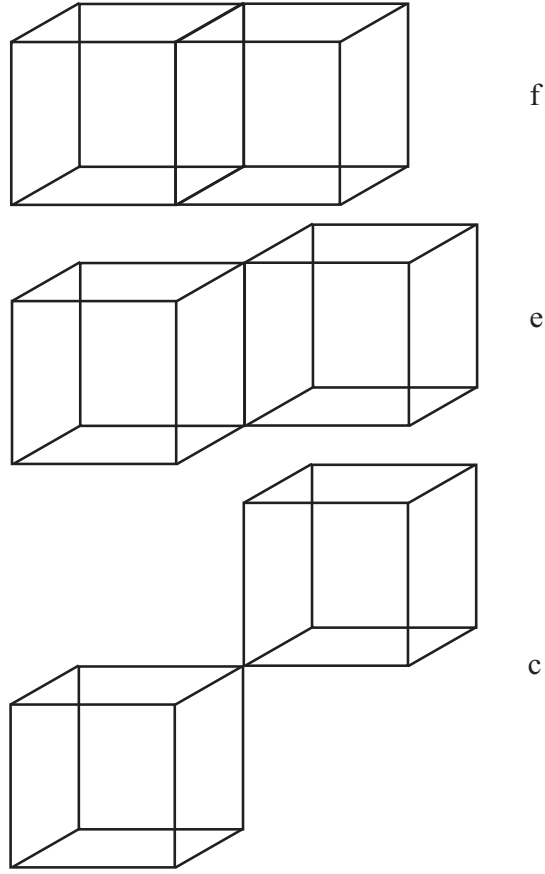}\vspace{1ex}
\caption{This figure illustrates the three cases of  touching
cubes for which the integrand is singular on a face, an edge, and
a corner, respectively.} \label{FIG1}
\end{figure}

We now introduce the symbols,
\begin{equation}\label{eq.sing.off.diag.integrals}
C_{l,1,1}^{l+1,1,1} \equiv  C_{\rm f}, \hskip 0.5cm C_{1,m,n}^{1,m+1,n+1} \equiv
C_{\rm e},\hskip 0.5cm C_{l,m,n}^{l+1,m+1,n+1} \equiv C_{\rm c},
\end{equation}for
the integrals between adjacent sub-cubes that have a face, an
edge, and a corner, in common, respectively (see Fig.\
\ref{FIG1}). These represent the terms in the sum
(\ref{eq.sum.algorithm}) that still are affected by the Coulomb
singularity. Using this notation formula (\ref{eq.sum.algorithm})
for the case $N=2$ gives,
\begin{equation}\label{eq.exact.expression}
C = C_{\rm f} +
C_{\rm e}+\frac{1}{3}C_{\rm c},
\end{equation}
which is the result (\ref{eq.result.for.the.four}) promised in the introduction.

The integrals (\ref{eq.sing.off.diag.integrals}) occur in the sum
(\ref{eq.I.sum.rewritten}) for $C$ the following number of times,
\begin{equation}\label{eq.numb.of.sing.off.diag.integrals}
N_{\rm f}=6N^2(N-1)
, \hskip 0.5cm N_{\rm e}= 12 N (N-1)^2, \hskip 0.5cm N_{\rm c}=
8(N-1)^3,
\end{equation}
respectively. Let us put,
\begin{eqnarray}\label{eq.numb.of.sing.off.diag.integrals.f}
F_{N} \equiv \frac{N_{\rm f}}{N^3(N^2-1)} = \frac{6}{N(N+1)} ,
\\
\label{eq.numb.of.sing.off.diag.integrals.e}
E_{N}\equiv\frac{N_{\rm e}}{N^3(N^2-1)}=\frac{12(N-1)}{N^2(N+1)},\\
\label{eq.numb.of.sing.off.diag.integrals.c} C_{N}\equiv  \frac{N_{\rm
c}}{N^3(N^2-1)}=\frac{8(N-1)^2}{N^3(N+1)},
\end{eqnarray}
and define the two quantities,
\begin{equation}\label{eq.def.deltaN}
\delta_N  \equiv  C_N^0 - F_{N}  - E_{N} \frac{1}{\sqrt{2}} -
C_{N}\frac{1}{\sqrt{3}},
\end{equation}
and, using this,
\begin{equation}\label{eq.def.IN1}
C_N^1 \equiv \delta_N + F_{N} C_{\rm f} + E_{N} C_{\rm e}+ C_{N}C_{\rm c}.
\end{equation}
Clearly $\delta_N$ is the sum of the terms in
(\ref{eq.brute.sum.I}) that approximate integrals that do not
contain singularities (in the interior or on the boundary). We
thus have that $\delta_2=0$ since for $N=2$ all the sub-cubes are
in contact. Therefore $C_N^1$ is an estimate of the integral $C$
by a sum in which the integrals containing surface singularities
have been replaced by their (unknown) exact values, while the
remaining ones are estimated by their inverse distance, Eq.\
(\ref{eq.basic.approximation}).

The function $C_N^1$ obeys both (since $\delta_2=0$),
\begin{equation}\label{eq.exact}
C_2^1 = C,
\end{equation}
and,
\begin{equation}\label{eq.limit2.for.I}
\lim_{N\rightarrow \infty} C_N^1 = C.
\end{equation}
Since the $\delta_N$ are known quantities the assumption that $C_N^1 = C$, in
equation (\ref{eq.def.IN1}), gives for each $N$ an equation in four unknowns ($C,
C_{\rm f}, C_{\rm e}, C_{\rm c}$). A system of four such equations,
\begin{equation}\label{eq.system}
C  - F_{N_k} C_{\rm f} - E_{N_k}
C_{\rm e} - C_{N_k} C_{\rm c} =  \delta_{N_k} \hskip 0.7cm k=1,2,3,4,
\end{equation}
can thus be solved for these unknowns. Now, each quadruple of
numbers $N_1, N_2, N_3, N_4$, will give us an estimate of the four
integrals. In calculating the $\delta_{N_k}$ the approximation
(\ref{eq.basic.approximation}) has only been used for integrals in
which the integrand does not become singular. Obviously one of the
numbers $N_k$ should always be chosen to be two since then one of
the equations of the system is exact.

\begin{table}\begin{tabular}{|l|l|l|l|l|l|l|l|}
\hline
$N_1$ & $N_2$ & $N_3$ & $N_4$ & $C^1$ & $C_{\rm f}^1$ & $C_{\rm e}^1$ &
$C_{\rm c}^1$\\
\hline $2$ & $3$ & $4$ & $5$ & 1.882304130 & 0.98272866 & 0.70632105 & 0.57976327  \\
\hline $2$ & $6$ & $8$ & $10$ & 1.882311519 & 0.98306698 & 0.70575406 & 0.58047142 \\
\hline $2$ & $11$ & $15$ & $19$ & 1.882312489 & 0.98340873 & 0.70521257 & 0.58107356 \\
\hline $2$ & $20$ & $25$ & $30$ & 1.882312615 & 0.98367876 & 0.70479560 & 0.58151474 \\
\hline $2$ & $30$ & $35$ & $40$ & 1.882312641 & 0.98390505 & 0.70445014 & 0.58187235 \\
\hline $2$ & $44$ & $50$ & $56$ & 1.882312647 & 0.98409569 & 0.70416088 & 0.58216823 \\
\hline
\end{tabular}
\caption[secondapprox]{\small \label{table.2nd.approx} The rows of this table
illustrate solutions of the system of equations (\ref{eq.system}). $C^1$ has
converged to $C$ in the last rows but the convergence to the other three integrals
is clearly slow.}
\end{table}

In Table \ref{table.2nd.approx} some results of this approach are shown. After
finding four different sets of quantities a standard linear equation solver delivers
four solutions to the linear set of equations. For $C$ this is clearly seen to give
excellent values. The three other integrals converge much more slowly but seem to
approach $C_{\rm f}\approx 0.984, C_{\rm e}\approx 0.704$, and $C_{\rm c} \approx
0.582$, respectively.

\section{Electrons in a homogeneous cube}
Here we will use crude estimates of the Hartree energy \cite{raimes} of electrons
that move in a cube of homogeneous positive charge density. Using this crude theory
we will investigate whether the electrons tend to delocalize in the cube or if a
state with localized electrons has lower energy.

We assume that the electrons either are delocalized in the cube and have constant
charge density in the cube or that they localize in one octant of the cube and have
constant charge density there. This means that we can treat either 8 electrons or 8
electron pairs. With these assumptions the electrostatic interaction energy can be
found from the results above. The kinetic energy is estimated essentially by means
of the uncertainty principle and the Pauli exclusion principle.

\begin{figure}
\centering \rotatebox{-90}{\includegraphics[width=200pt,height=300pt]{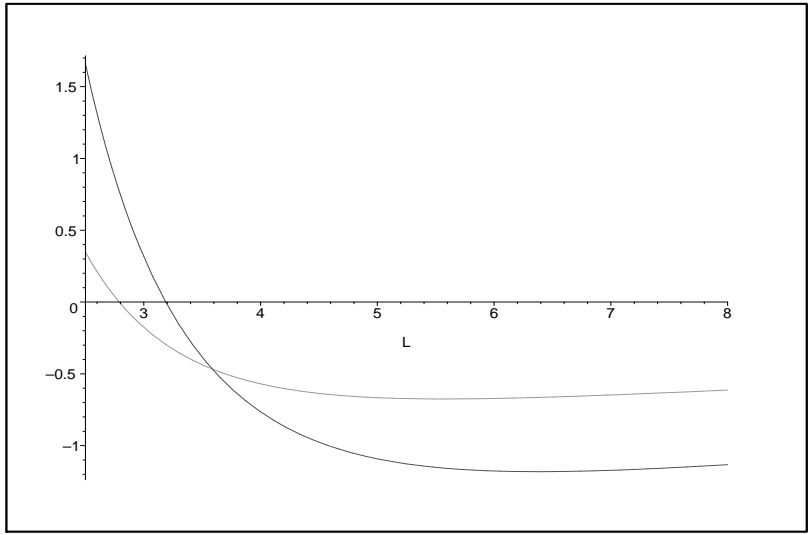}}
\protect\caption {The energies of eight electrons in a cube of side $L$ as function
of $L$ for the localized and delocalized cases, respectively. For $L>3.59$ the
localized state has lower energy. } \label{FIG8}
\end{figure}

We start with 8 electrons in delocalized states. They are assumed to move in a cube
of side $L$ and positive charge $8e$. The energy is then the sum of the kinetic
energy,
\begin{equation}\label{eq.T.8.deloc}
T_{8d} =
\frac{\hbar^2}{2m}\left[2\frac{3}{L^2}+6\left(\frac{2}{L^2}+\frac{1}{(L/2)^2}
\right) \right] ,
\end{equation}
and the electrostatic energy,
\begin{equation}\label{eq.V.8.deloc}
V_{8d} = \frac{1}{2}\frac{(8e)^2}{L}C -8\frac{8e^2}{L}C + \frac{8\cdot 7}{2}
\frac{e^2}{L}C.
\end{equation}
In the kinetic energy the two first electrons are assumed delocalized over the cube
without nodes in the wave function. The remaining six must then go into the three
degenerate states with one node. The first term in the electrostatic energy is the
self energy of the positive background. Then follows the attraction between the
background and the eight delocalized electrons. The final term is the sum of the
$8\cdot 7/2$ electron-electron pair repulsion terms. Simplifying this gives the
total energy
\begin{equation}\label{eq.E.8.deloc}
E_{8d} = \frac{\hbar^2}{2m} \frac{42}{L^2} - C e^2 \frac{4}{L}.
\end{equation}
The energy of this closed shell delocalized state should now be compared to the
energy of the ferromagnetic localized state with the electrons in one corner each.
We find,
\begin{equation}\label{eq.T.8.loc}
T_{8l} = \frac{\hbar^2}{2m} 8\frac{3}{(L/2)^2} = \frac{\hbar^2}{2m}\frac{96}{L^2} ,
\end{equation}
for the kinetic energy since all 8 electrons now sit in cubes (octants) of side
$L/2$. They are however alone in their corners (octants) so the Pauli principle is
automatically obeyed. The electrostatic energy becomes
\begin{equation}\label{eq.V.8.loc}
V_{8l} = \frac{1}{2}\frac{(8e)^2}{L}C -8\frac{e^2}{L/2}\left(C + 3C_{\rm f}+3C_{\rm
e}+ C_{\rm c} \right) + \frac{8}{2}\frac{e^2}{L} \left( 3C_{\rm f}+3C_{\rm e}+
C_{\rm c} \right).
\end{equation}
Simplification of this using Eq.\ (\ref{eq.exact.expression}) gives the total energy
\begin{equation}\label{eq.E.8.deloc}
E_{8l} = \frac{\hbar^2}{2m} \frac{96}{L^2} - C e^2 \frac{8}{L}.
\end{equation}
If we introduce atomic units ($\hbar = e= m=1$) so that length is measured in units
of the Bohr radius we can plot the two energy curves,
\begin{eqnarray}
  E_{8d} &=&   \frac{21}{L^2} - C  \frac{4}{L}, \\
  E_{8l} &=&   \frac{48}{L^2} - C  \frac{8}{L},
\end{eqnarray}
and get the results of Fig.\ \ref{FIG8}.

\begin{figure}
\centering \rotatebox{-90}{\includegraphics[width=200pt,height=300pt]{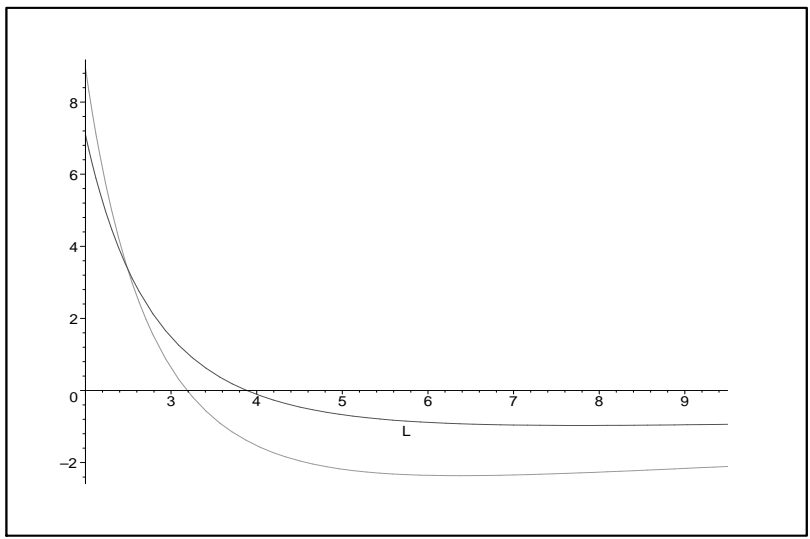}}
\protect\caption {The energies of eight electron pairs in a cube of side $L$ as
function of $L$ for the localized and delocalized cases, respectively. For $L>2.49$
the localized state has lower energy. } \label{FIG16}
\end{figure}

Finally we give the corresponding results for 16 electrons sharing orbitals
pairwise. In the electrostatic energy one can then essentially change the particle
charge $e$ to $2e$ and add the contributions from the repulsion within the pairs.
This gives the two curves,
\begin{eqnarray}
  E_{16d} &=&   \frac{117}{2 L^2} - C  \frac{8}{L}, \\
  E_{16l} &=&   \frac{192}{2 L^2} - C  \frac{16}{L}.
\end{eqnarray}
for the delocalized and localized energies respectively. These
curves are plotted in Fig.\ \ref{FIG16}.

One notes that the localized states always have lower electrostatic energy simply
because in these states the electrons are better at avoiding each other. For small
$L$-values the delocalized states always have lower energy because of the
uncertainty principle. The curves in these plots resemble those of Wigner
\cite{wigner34,wigner38} who predicted that localization gives lower energy in
metals at low densities. This phenomenon is called Wigner crystallization.

\section{Conclusions}
I am not aware of any comparable study of the electrostatic interaction energies of
homogeneous cubic charge distributions. The algebraic and combinatoric tricks used
to eliminate the Coulomb singularities in the integrals seem partly new, as well as
the result of Eq.\ (\ref{eq.result.for.the.four}). It is possible that these ideas
can be generalized to more general integration problems involving the Coulomb
singularity. It is a further bonus that these insights into the electrostatics of
cubes and their sub-cubes can be used to make simple estimates for the Hartree
energy of electrons distributed in cubes in different ways. Such simple model
systems are of value for the qualitative understanding of more complex systems.

\end{document}